\documentclass{elsart3p}
\usepackage{epsfig}
\usepackage{amsmath}
\usepackage{sidecap}
\usepackage{color}

\newcommand{\MDcomm}[1]{{\color{black}#1}}
\newcommand{\escomm}[1]{{\color{black}#1}}

\begin{document}
\begin{frontmatter}
\title{Dipole excitations of Ar substrate in contact with Na clusters} 

\author{P.~M.~Dinh\corauthref{cor}$^a$}
\author{, P.-G.~Reinhard$^b$, E.~Suraud$^a$}

\corauth[cor]{Corresponding author\\{\it Email-address}~:
  dinh@irsamc.ups-tlse.fr} 
\address{$^a$Laboratoire de Physique Th\'eorique, Universit\'e Paul
  Sabatier, CNRS, F-31062 Toulouse C\'edex, France}
\address{$^b$Institut f{\"u}r Theoretische Physik, Universit{\"a}t
  Erlangen, D-91058 Erlangen, Germany}


\begin{abstract}
We analyze the excitation of Ar substrate in contact with Na clusters
using a previously developed hierarchical model for the description of
the system cnstituted  of a highly reactive metal cluster in contact
with a rather 
inert substrate. Particular attention is paid to the dipole excitation
of the Ar atoms and the energy stored therein. The Na clusters are
considered at different charge states, anions, cations, and neutral
clusters for the case of deposition and a highly ionized cluster
embedded in a matrix. It is found that the dipole polarization of the
Ar atoms stores the largest fraction of energy in the case of charged
clusters. Some, although smaller, polarization is also observed
for polar clusters, as Na$_6$. 
The effect is predominantly induced by the electrostatic interaction.
\end{abstract}

\end{frontmatter}

\section{Introduction}

The study of clusters in contact with an environment has motivated
many investigations over the years. Two major situations can be
considered, namely a cluster deposited on a surface or a cluster
embedded inside a matrix. Both situations bear some similarities
and bring complementing information.  Clusters on surfaces
provide interesting perspectives for basic research and for
applications to nano-structured materials \cite{Hab94b,Bin01},
e.g.,  the synthesis of deposited clusters, either
controlled growth of elementary units on a surface by molecular beam
epitaxy \cite{Bru00} or direct deposition of size-selected clusters on
a substrate \cite{Har00}. One should also mention the non-destructive
deposition technique of metal clusters on metal surfaces using a thin
rare gas film above the metal surface \cite{Lau05}.  At the side of
embedded species, one can refer to the many studies exploring the
optical response in various dynamical scenarios
\cite{Har93,Gau01,Die02} complementing similar investigations in
the case of free clusters.  Note furthermore that embedded clusters
(or molecules) may also be viewed as model systems for a detailed
analysis of radiation effects in matter \cite{Bar02,Niv00}.

>From the theory side, the description of clusters in contact with an
environment implies an extra complexity because one needs to
account for the degrees of freedom of the substrate.  Theoretical
descriptions thus predominantly employ classical molecular dynamics
with effective atom-atom forces, see \cite{Xir02}.  This was, for
example, done for the deposition dynamics of Cu clusters on
metal~\cite{Che94} or Ar~\cite{Rat99} surfaces, and of Al or Au
clusters on SiO$_2$~\cite{Tak01a}. Such simplified approaches overlook
possible effects from electronic degrees of freedom which can become
crucial in metal clusters, particularly if a finite net charge is
involved. It is thus often necessary to account explicitely
for electronic degrees of freedom. Fully detailed calculations
have been performed which treat all constituents with
their electronic dynamics, e.g. for the structure of small Na
clusters on NaCl~\cite{Hak96b} or the deposit dynamics of Pd clusters
on a MgO substrate~\cite{Mos02a}. But the numerical effort of such
detailed computations quickly grows huge. Furthermore, these subtle
models are hardly extendable to truly dynamical situations, to larger
clusters or substrates, and to systematic explorations for broad
variations of conditions.  This leaves space for a manifold of
approximations allowing an affordable compromise between reliability
and expense. Such methods are often called
quantum-mechanical-molecular-mechanical (QM/MM) models and have been
applied for instance to chromophores in
bio-molecules~\cite{Gre96a,Tap07}, surface
physics~\cite{Nas01a,Inn06}, materials
physics~\cite{Rub93,Kur96,Ler98,Ler00}, embedded
molecules~\cite{Sul05a} and ion channels of cell
membranes~\cite{Buc06}. We have developed such a QM/MM
modeling, primarily in the case of Na in contact with Ar
\cite{Dup96,Ger04b,Feh05a} and successfully applied this method to
deposition dynamics on finite Ar clusters~\cite{Din07a} or on Ar
surfaces~\cite{Din07b}, as well as to irradiation scenarios in the
case of embedded clusters both in the linear (optical response
\cite{Feh07a}) and non linear (hindered explosion
\cite{Feh07b,Feh07c}) domains.
More recently, we have  extended the modeling to a MgO substrate
\cite{Bae07a}.
The originality of this approach lies in the fact that the environment
polarizability is treated dynamically, a key aspect especially when
charged species are considered \cite{Din08}.

In this paper, we want to further investigate the
importance of this dynamical treatment of the environment's
polarizability by exploring how the substrate's dipoles 
respond to a perturbation. We shall consider two typical scenarios~:
Cluster deposition on a surface and irradiation of an embedded
cluster. In both cases, we shall explore in particular the impact of
charge, either because the deposited species is charged, or because
the embedded cluster acts as a chromophore in a laser field and thus
quickly acquires charge after irradiation. This exploration is focused
on theoretical aspects but is qualitatively closely related to recent
experiments in which the deposition of charged Ag clusters on a Ar
matrix, itself deposited on a Au surface has been studied
\cite{Sie06,Har07}.
It was shown in this experiment that the substrate, in spite of its a
priori inert character, acquires a substantial inner excitation which 
plays a key role in the whole process. Exploring the dipole degrees of 
freedom in our Na-Ar combination then represents the simplest model case in 
relation to these experiments. We shall see that indeed, as suggested by 
our earlier investigations \cite{Din08}, these internal degrees of freedom 
are readily excited in most of the studied cases. 

The paper is organized as 
follows. After a brief reminder of the content of the model used here,
we quickly focus on relevant test cases. We first show the overall
importance of charge effects and then study the spatial extension of
the inner excitation of the environment. We take examples from
deposition processes and irradiation of embedded clusters.

\section{Model}
\label{sec:model}

We first give  a very brief summary of the hierarchical description of
the combined Na--Ar system. We treat the metal atoms in full
microscopic detail at the level of Time Dependent Local Density
Approximation (TDLDA) for the valence electrons of the Na cluster. 
An average self-interaction correction is applied to LDA in
order to put the ionization potential of the Na system at the correct
place \cite{Leg02}. 
Electronic degrees of freedom are coupled to Molecular Dynamics (MD)
for the ions. Details on this very successful TDLDA-MD approach for
free clusters can be found in~\cite{Cal00,Rei03a}.  The environment
(substrate or matrix) consists out of Ar atoms which are
described by classical degrees of freedom, both in terms of position
and dipole moment. The latter serves to take into account the
dynamical polarizability of the substrate atoms.
They have been adjusted carefully to recover the static and
dynamical polarizabilities of Ar atoms. Electronic emission is not
possible at the side of Ar atoms which limits the violence of
the processes studied. But it is easy to check, in terms of the
amplitude of Ar dipoles, how much energy is actually absorbed by an Ar
atom.  
More precisely, this internal excitation energy of the Ar
atoms is related, in our model, to the Ar dipole amplitude $\mathbf
d$ by 
\begin{equation}
E_{\rm dip} = \frac{1}{2} \, e^2 \, \frac{{Q_{\rm Ar}}^2}{\alpha_{\rm
    Ar}} \, \mathbf d^2,
\label{eq:Edip}
\end{equation}
where $\alpha_{\rm Ar}$ is the static polarizability of bulk Ar, and
$Q_{\rm Ar}$ the effective charge of the Ar cores \cite{Feh05a}.

As long as this energy (and correspondigly the dipole amplitude)
remains safely below the Ar atom ionization potential, the model is
perfectly applicable. We checked that this is always the case in all
situations encountered here. For then, the dipole amplitude will
represent the only possible inner excitation of Ar atoms.  The Ar
atoms are coupled to the Na by a long range polarization
potential and some short range repulsion to
account for the Pauli blocking of cluster electrons in the vicinity of
the Ar cores. The model has been calibrated to measured properties of
typical Na-Ar systems. We refer the reader
to~\cite{Dup96,Ger04b,Feh05b} for a detailed description of the model
and the detailed fitting of the various parameters.

In the case of deposition dynamics, the Ar(001) surface is 
modelled through an Ar$_{384}$ system, consisting in six layers of 8$\times$8
Ar atoms with atoms in the two lowest layers frozen at bulk crystal
positions. The layers are furthermore periodically
repeated in both lateral directions, thus simulating bulk material in
these two dimensions. We have checked that the finiteness of the
sampling does not alter the results, at least qualitatively
\cite{Din07a,Din08}. The dynamics is then initialized by placing the
projectile  at a finite distance (15 a$_0$) from the surface and
boosting it with a given initial kinetic energy $E_0$, towards the
substrate and along the direction normal to it (denoted by $z$ in the
following).
%
%
In the case of embedded species, the system is constructed by
considering first a finite piece of bulk Ar, drilling a small hole in
the center to insert the Na$_8$ cluster and then optimizing the whole
ensemble (cluster + embedding material Ar$_{434}$) at the side of
cluster electrons and ions and in terms of Ar positions and dipoles. The
system is then irradiated by a laser and its response followed in
time.  We analyze the subsequent dynamics in terms of detailed ionic
and atomic coordinates and dipoles.

\section{Impact of charge}

Let us first consider a typical example, namely the deposition of a
Na$_6^+$ cluster (consisting in a pentagon with a top ion on its
symmetry axis) on Ar(001). We take a charged cluster in the spirit of
\cite{Sie06,Har07} and a moderate initial kinetic energy in order to
observe sticking of the cluster on the surface with no destruction of
the substrate \cite{Din07b,Din07c}.
\begin{figure}[htbp]
\begin{center}
\includegraphics[width=\linewidth]{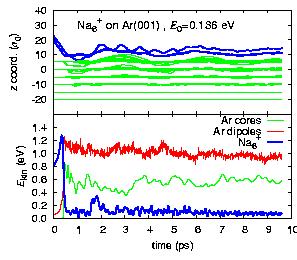}
\caption{Time evolution of coordinates and energies during
  dynamical deposition of Na$_6^+$ on Ar(001) with initial
  kinetic energy $E_0=0.136$ eV/ion. 
  Upper panel: $z$ coordinates.
  Lower panel: total kinetic energy of Ar cores, Na
  ions and total dipole excitation energy.}
\label{fig:na6+}
\end{center}
\end{figure}
Fig.~\ref{fig:na6+} shows the time evolution of positions and
energies. In the upper panel are plotted the $z$-coordinates of ions
(Na$_6^+$) and surface (Ar) atoms. The $z$ direction is the
direction along the initial cluster velocity, perpendicular to the
surface. The Na$_6^+$ cluster is significantly perturbed as
one can see from the large oscillations of the initially top ion going
through the pentagon plane.  Nonetheless, the overall process
converges steadily to a robust sticking of the cluster on the
surface. The surface itself is perturbed with some ionic
rearrangement but with preservation of layer structure. The lower
panel of Fig.~\ref{fig:na6+} displays the time evolution of energies,
i.e. kinetic energies of cluster ions and Ar atoms, and the energy
stored in Ar dipoles, see Eq.~(\ref{eq:Edip}).
The time evolution is rather simple with an almost instantaneous
transfer of cluster kinetic energy to substrate degrees of freedom
\cite{Din07a,Din07b}. 
Note that some part of initial energy is flowing into potential energy
which is not shown here. 
Still, the interesting feature is that, while Ar atoms
acquire a significant kinetic energy of about half the maximum one of
the cluster, they store about twice as much energy in their
dipoles. This is a key aspect. It means that the substrate is not only
heated up by cluster impact but also internally excited at the
side of each constituent atom.
This shows that a proper treatment of the dynamical surface
polarizability can be crucial, particularly if a charge is involved.
The energy sharing is established almost instantaneously at impact
time and energies then remain rather constant in time.

\begin{figure}[htbp]
\begin{center}
\includegraphics[width=0.9\linewidth]{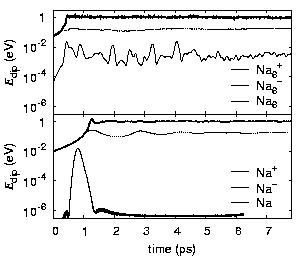}
\caption{Total excitation energy of the dipoles in the substrate
  Ar$_{384}$, after deposition of various metal species, as indicated,
  with initial kinetic energy $E_0=0.136$ eV/ion, as a function of time.}
\label{fig:depos_edip}
\end{center}
\end{figure}

The strong effect at the side of dipoles observed in Fig.~\ref{fig:na6+} has to be explored
further in order to try to identify where it comes from. Using a
Na$_6^+$ cluster as a projectile implies two possible effects, from
charge and/or mass. We have considered here the case of a cationic
cluster but an anionic one could be envisioned as well. We thus try to
disentangle charge and mass effects by considering various possible
combinations namely Na$_6$, Na$_6^+$, Na$_6^-$, Na, Na$^+$ and Na$^-$.
We hence focus on dipole energies of the six test cases, which are
plotted as a function of time in Fig.~\ref{fig:depos_edip}.
As compared to the Na$_6^+$ case, 
the dipole energies are suppressed by several
orders of magnitude (2 for Na$_6$ and even 6 for Na)  for
neutral projectiles, while with
charged projectiles, they are of the same order, with even a higher
dipole energy for a positive charge by a factor about 1.5 with respect
to the Ar core kinetic energies (not shown).  
The dominant effect is obviously due to charge. In both mass cases (Na
vs Na$_6$), the dipole energies associated to charged projectiles are
several orders of magnitude larger than for the neutral
cases. For neutral systems, one can spot a tiny mass effect
by comparing Na to Na$_6$ but which, however, remains negligible with
respect to charge effects. One also can notice a difference between
anions and cations, the difference lying within one order of
magnitude. The negatively charged electron cloud of the anions
experience a stronger Pauli repulsion from the Ar cores.
As a consequence, these projectiles cannot transfer as
much energy to the Ar substrate as in the positively charged
cases. This confirms earlier calculations on similar
systems~\cite{Din08}.

\section{Details of energetics}

From the previous section, we learnt that the matrix is qualitatively
excited the same way either by Na$^+$ or Na$_6^+$. To get more insight
into the energetics, we present in this section the detail of the
energy sharing for the case of deposition of Na$^+$, which kicks out
the Ar atom just below the impact point and finally stays between the
first and the second layers of the surface~\cite{Din08}. This case has
the advantage that no valence electron contributes and that only one
Na$^+$ ion is involved. Thus, there remain only five components to the
total energy $E_{\rm tot}$, namely
\begin{equation}
\label{eq:Etot}
  E_{\rm tot} 
  = 
  E_{\rm pot}^{\rm mat} + E_{\rm dip} + E_{\rm coupl} +
  E_{\rm kin}^{\rm Na^+} + E_{\rm kin}^{\rm mat} 
  \quad.
\end{equation}
The kinetic energies are obvious. The contributions to the potential
energy are~: $E_{\rm pot}^{\rm mat}$ as the potential energy of the Ar
matrix, consisting out of the Coulomb energy and a contribution from
the Ar-Ar core repulsion, $E_{\rm dip}$ as the dipole
excitation energy defined in Eq.~(\ref{eq:Edip}), and $E_{\rm coupl}$
for the coupling energy between the Na$^+$ and the Ar matrix.  As it
should be, the total energy is well conserved during the deposition
process \MDcomm{(up to a relative error less than 0.01\% over the
whole simulation)}. 
\begin{figure}[htbp]
\begin{center}
\includegraphics[width=0.9\linewidth]{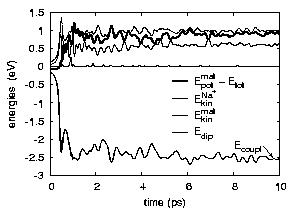}
\caption{Time evolution of the various contributions to the total energy 
as given in eq. (\ref{eq:Etot}) in the deposition
  of Na$^+$ on Ar$_{384}$, with respect to the total initial energy. 
}
\label{fig:energy}
\end{center}
\end{figure}
Fig.~\ref{fig:energy} displays various terms of
Eq.~(\ref{eq:Etot})
\escomm{as a function of time. At initial time, the total energy 
$E_\mathrm{tot}$ of course consists of the initial Na$^+$ kinetic
energy but mostly of the matrix potential energy $E_{\rm
  pot}^{\rm mat}$ (99.8~\%) with small additional contributions from
the coupling and dipole energies. These two latter contributions do
not exactly vanish at initial time due to the initial finite distance
at which the Na$^+$ ion is placed with respect to the Ar
surface. Since the initial potential energy surface scales
with the matrix size (hence with the finiteness of the representation
of the surface), it makes little sense to keep it in the
picture. We should hence  
remove it from the total energy (it is anyway a constant). But for
sake of readibilty of Fig.~\ref{fig:energy}, it appears simpler to
remove the total initial energy $E_\mathrm{tot}(t=0)$ so that the
components of the energy now sum up to zero at any time, as is clear
form the figure.
} 
All contributions have a similar time structure. There is a fast change
within the first ps and then the values stabilize with some final
fluctuations.  The final share is the following~: The largest and
attractive contribution comes from the coupling energy. This is
counterweighted to \escomm{comparable} parts between 
matrix potential energy, matrix kinetic energy, and dipole energy
\escomm{(still with the latter taking the lead)} .
The kinetic energy of the deposited Na$^+$ ion is negligible as it
should be for a well bound particle.

\escomm{It is also interesting to}
show the share of energies in relative units, we compare it with
the total attachment energy of Na$^+$ to the surface which is
$E_\mathrm{deposit}=-4.7$ eV. Relative to the absolute value of that
energy, we have the contributions~:
\begin{center}
\begin{tabular}{ccccc}
\multicolumn{1}{c}{$\;\; E_\mathrm{pot}^\mathrm{mat}-E_\mathrm{tot}\;\;$}
&
\multicolumn{1}{c}{$\;\; E_\mathrm{coupl}\;\;$}
&
\multicolumn{1}{c}{$\ \  E_\mathrm{dip}\ \ $}
&
\multicolumn{1}{c}{$\quad E_\mathrm{kin}^\mathrm{Ar}\quad $}
&
\multicolumn{1}{c}{$\quad E_\mathrm{kin}^\mathrm{Na}\quad $}
\\
  64\%
&
  56\%
&
 21.9\%
&
 13.7\%
&
 0.055\%
\\
\end{tabular}
\end{center}
%
%
The main contributions thus come from the potential energy due the
rearrangement of the whole matrix, from the strong coupling between
the Na$^+$ and the matrix (since the metal ion finally locates between
the two first Ar layers), and from the internal excitation of each Ar
atom. 

\section{Localization of the dipoles}

\begin{figure}[htbp]
\begin{center}
\includegraphics[width=\linewidth]{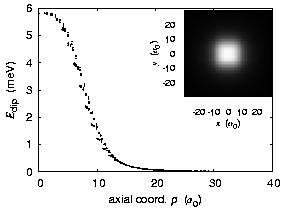}
\caption{
Axial distribution of the dipole energies in the first layer
of the Ar(001) substrate at impact time, for the deposition of
Na$^+$ with initial kinetic energy $E_0=0.136$ eV. The insert shows
a top view of the distribution in the first layer ($x-y$ plane).
}
\label{fig:nap_hotspot}
\end{center}
\end{figure}

The role of charge in the dipole energy suggests to explore in more
detail the spatial distribution of the dipole excitation,
with the intuition that it might resemble the response to a mere
charge.
We first present in Fig.~\ref{fig:nap_hotspot} the dipole
energies as a function of the axial coordinate $\rho=\sqrt{x²+y²}$ of
the Ar cores, at impact time and only in the upper layer of the Ar
substrate, for the case of the deposition of Na$^+$ with initial
kinetic energy of 0.136 eV.
The impact point is located at $\rho=0$. We observe a high excitation
of the dipoles which is strongly located around the impact point. 
As we shall see below, one does not observe any sizable evolution 
of this distribution,
at least up to
the times computed here, in terms of total dipole energies (see
Figs.~\ref{fig:na6+} and \ref{fig:depos_edip}).
\begin{figure*}[htbp]
\begin{center}
\includegraphics[width=0.9\linewidth]{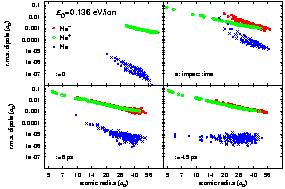}
\caption{Root mean square dipole moments of Ar atoms as function of Ar atomic
  radius, at four different times as indicated, for deposition with
  initial kinetic energy $E_0=0.136$ eV of Na, Na$^+$ and Na$^-$ on
  Ar(001).} 
\label{fig:na_distrib}
\end{center}
\end{figure*}
\begin{figure*}[htbp]
\begin{center}
\includegraphics[width=0.9\linewidth]{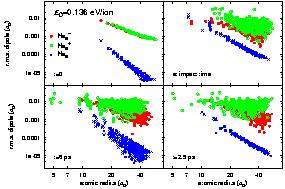}
\caption{Root mean square dipole moments of Ar atoms as function of Ar atomic
  radius, at four different times as indicated, for deposition with
  initial kinetic energy $E_0=0.136$ eV/ion of Na$_6$, Na$_6^+$ and
  Na$_6^-$ on Ar(001).} 
\label{fig:na6_distrib}
\end{center}
\end{figure*}
In order to analyze this aspect in more detail,  actual dipole moments (in fact, r.m.s. dipole
moments $\bar{d}=\sqrt{d_x^2+d_y^2+d_z^2}$) as a function of time are
plotted in Figs~\ref{fig:na_distrib} (case of Na monomers) and
\ref{fig:na6_distrib} (case of Na$_6$ clusters).  We discuss
both cases simultaneously because they deliver very similar messages
and because we have seen that mass effect is expectedly small. And
yet, it is interesting to countercheck by comparing both, since the
finite extension of Na$_6$ might influence the spatial distribution of
Ar dipole excitations. In both figures are plotted four
snapshots corresponding to initial time, time of first
impact and two later instants. 
There is obviously no significant time evolution of the
distributions. One rather
observes in Figs.~\ref{fig:na_distrib} and \ref{fig:na6_distrib} very
similar pattern in both late times and both systems. One can spot
differences in the neutral case; however the corresponding values are
very small and one can probably ignore the point. Initial
distributions as well are very similar in both figures, at least for
charged clusters/atoms. The very regular trend is typical of a
population of dipoles subject to a distant point charge, as is the
case for initial states. The most interesting panels are probably the
ones corresponding to impact time~: One can note a significant
perturbation of the distributions. This is especially true in the case
of Na$_6$, most probably because of its finite extension which
perturbs the matrix on a larger range. The effect actually remains for
longer times, with a somewhat (although not significantly)
fuzzier distribution in the Na$_6$ case.
The predominant effect is that finally the dipole
excitations remain strongly located close to the cluster impact
point. Everything thus looks as if the charge was simply put closer to
the surface with the ensuing enhanced response of the dipole
accompanied by a bit of ``noise'' at the distant points.
In that respect, the details of the deposition dynamics are to a large
extent irrelevant. Only charge and its localization are really
important.  This is in accordance with our previous study of neutral
or charged Na monomer deposition/reflection on an Ar
surface~\cite{Din08}. Indeed, as soon as the Na projectile is
reflected, no sizable dipole energy beyond ``noise'' is left
in the Ar substrate. Substantial dipole response is seen
only at impact time, that is, when the (charged) projectile is
sufficiently close to the Ar atoms. Thus, in these deposition
processes, we are facing predominantly an electrostatic
effect of the polarization of the surface. This effect is, however,
energetically important and needs to be taken into account.

\section{An example from embedded clusters}

We have focused up to now on the case of cluster deposition. It is
also interesting to study what occurs in the embedded case
in which high charge states can be easily attained by laser
excitation. We take as an example the irradiation of a Na$_8$ cluster
embedded inside a finite Ar$_{434}$ matrix. The laser pulse
is kept short to concentrate the ionization to a fairly well defined
initial time and the intensity is tuned such that the irradiation
leaves the cluster with a net $3+$ charge. A Coulomb
explosion of the cluster is hindered by the matrix which stabilizes
that high charge state, but allows a sizable oblate expansion of the
cluster~\cite{Feh07b,Feh07c}.
\begin{figure}[htbp]
\begin{center}
\includegraphics[width=0.9\linewidth]{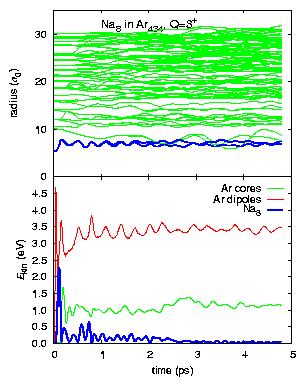}
\caption{
  Time evolution of radial coordinates 
  $r=\sqrt{x^2+y^2+z^2}$ (upper panel) and kinetic energies
  (lower panel) for 
  Na$_8$ excited by laser of intensity
  $2\times 10^{12}$ W/cm$^2$, frequency $\omega=1.9$ eV, and 
  pulse length with full width at half maximum of 33 fs.
  The laser lifts the Na$_8$ within less than 100 fs into
  a charge state of $Q=3^+$.
} 
\label{fig:na8}
\end{center}
\end{figure}
The point is illustrated in the upper panel of Fig.~\ref{fig:na8}
which shows the ionic and atomic positions along laser polarization
axis as a function of time. The Ar matrix is also strongly perturbed
with significant atomic rearrangement but the whole system finally
remains stable. The lower panel of Fig.~\ref{fig:na8} displays
energies~: The kinetic energy of Na ions and Ar cores, and
the energy stored in the Ar dipoles. The energy balance is to some
extent incomplete as it misses the potential energy of cluster and
matrix.  But, as in case of deposition, the kinetic and dipole
energies provide sufficient information about energy flow. We see in
the lower part of Fig.~\ref{fig:na8} a behavior similar to the
deposition case, namely that the initial Na motion is quickly damped
and its energy accordingly transferred to the matrix. Again, one
observes a significant amount of energy stored in the Ar dipoles,
about three times as much as in the Ar core kinetic energy.

\begin{figure*}[htbp]
\begin{center}
\includegraphics[width=0.9\linewidth]{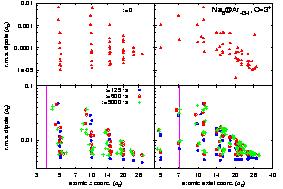}
\caption{Root mean square Ar dipoles for the hindered Coulomb explosion
  of Na$_8$ embedded in Ar$_{434}$, exposed to a laser of intensity
  $2\times 10^{12}$ W/cm$^2$, frequency $\omega=1.9$ eV, and FWHM=33
  fs. Left panel~: Distribution as a function of the Ar $z$
  coordinates; right panels~: That as a function of the Ar axial
  coordinates $\rho=\sqrt{x^2+y^2}$. Top panels~: At initial time;
  bottom panels~: For three subsequent times as indicated. The maximum
  of excitation energy observed in the bottom right panel at 11 $a_0$
  is due to the oblate deformation of the created Na$_8^{3+}$. The
  vertical lines in the bottom panels indicate the corresponding
  coordinates of the outer Na$_8$ ions.}
\label{fig:na8_distrib}
\end{center}
\end{figure*}
To complement the analysis of this irradiation case, we consider again
the spatial distributions of dipoles at various instants in
Fig.~\ref{fig:na8_distrib}. The geometry is now a bit more complicated
and we should care for both directions, along the laser polarization
($z$ direction) and perpendicular (axial direction) to it. We thus
slightly change the representation of the distribution in
Fig.~\ref{fig:na8_distrib} considering separately both directions.
The situation is also different from the charged deposition case to
the extent that the cluster is initially neutral. This explains the
rather "democratic" distribution of dipoles all over the matrix at
initial time (upper panels). This confirms earlier findings on the
role of Ar polarizabilities even in low energy phenomena such as
optical response \cite{Feh05a,Feh07a}. The initial dipole
amplitudes are rather small as compared to the values which they
acquire when the cluster becomes charged (mind the scales of upper
and lower panels).  The lower panels of Fig.~\ref{fig:na8_distrib}
show large values of dipoles, comparable to deposition of charge
clusters (see Figs.~\ref{fig:na_distrib} and
\ref{fig:na6_distrib}). The distribution, though, looks at first
glance quite different from the deposition cases with a maximum at a
finite distance from center rather (in radial coordinates, right lower
panel) than a monotonously decreasing shape. The effect is in fact due
to the finite (large) extension of the Na cluster after
irradiation. Indeed, a sizable fraction of Ar sites are embedded in
the Na cluster electron cloud and thus see a
screened charge, whence the reduced dipole polarization. In order to
exemplify the point, we have also plotted in
Fig.~\ref{fig:na8_distrib} the actual position of the most external Na
ions and this makes the pattern clear, with an effect only along
radial coordinate due to the oblate shape of the irradiated cluster
\cite{Feh07b,Feh07c}.  Apart from that detail, we find again a
dominantly static charge effect explaining the dipole polarizations of
Ar atoms, much similar to the deposition case with localization
prevailing again and little own dynamics.


\section{Conclusion}

We have discussed in this paper the atomic response of an Ar
surface or matrix perturbed by a metal cluster either through
deposition or irradiation by a laser. We have focused the analysis on
the internal response of the Ar atoms by studying their single
internal degree of freedom in our model, namely their dipole
polarizability. We have seen that the dominant effect is due to
cluster charge. Mass effects are mostly negligible with respect to
charge effects. We have also analyzed how the dipole excitations are
spatially distributed. We have observed that the excitation remains
strongly localized and essentially does not evolve in time. It
resembles in many respect the effect of a finite charge deposited at
some place in the system. This holds true, up to details, for all
situations involving charges, in deposition as well as in irradiation
dynamics.  Independent dynamical evolution of the Ar dipoles
creates some background noise which, however, remains quantitatively
unimportant at the present level of analysis.  The quantitatively very
important effect remains the nearly static dipole deformation.  The Ar
dipoles store a sizable fraction of the available energy.
This is in perfect qualitative agreement with
recent experiments on similar systems.

\bigskip

Acknowledgments~: This work was supported by the DFG, project nr. RE
322/10-1, the French-German exchange program PROCOPE nr. 07523TE, the
CNRS Programme ``Mat\'eriaux'' (CPR-ISMIR), Institut Universitaire de
France, the Humboldt foundation, a Gay-Lussac price, and the French
computational facilities CalMip (Calcul en Midi-Pyr\'en\'ees), IDRIS and
CINES.

\bibliographystyle{apsrev}

\end{document}